\documentclass{JINST}
\usepackage{graphicx}

\newcommand{\ANKARA}        {1}
\newcommand{\ANNECY}        {2}
\newcommand{\ASSERGILNGS}   {3} 
\newcommand{\BARI}          {4}
\newcommand{\BARIINFN}     {5}
\newcommand{\BERN}          {6}
\newcommand{\BOLOGNAINFN}  {7}
\newcommand{\BOLOGNA}       {8}
\newcommand{\BRUSSELS}      {9}
\newcommand{\DUBNA}         {10}
\newcommand{\FRASCATI}      {11}
\newcommand{\FUNABASHI}     {12}	
\newcommand{\HAMBURG}       {13}
\newcommand{\GAZWADONG}     {14}
\newcommand{\KARIYA}        {15} 
\newcommand{\KOBE}          {16}
\newcommand{\LAQUILA}       {17}
\newcommand{\LYON}          {18}
\newcommand{\MOSCOWINR}     {19}
\newcommand{\MOSCOWITEP}    {20}
\newcommand{\MOSCOWLPI}     {21}
\newcommand{\MOSCOWSINP}    {22}
\newcommand{\NAGOYA}        {23}
\newcommand{\NAPOLIINFN}   {24}
\newcommand{\NAPOLI}        {25}
\newcommand{\NEUCHATEL}     {26}
\newcommand{\OBNINSK}       {27}
\newcommand{\PADOVAINFN}   {28}
\newcommand{\PADOVA}        {29}
\newcommand{\ROMA}          {30}
\newcommand{\SALERNO}       {31}
\newcommand{\STRASBOURG}    {32}
\newcommand{\URBINO}        {33}
 \newcommand{\UTSUNOMIYA} {34}
\newcommand{\ZAGREB}        {35}
\newcommand{\ZURICH}        {36}

\newcommand{\OperaInstitutes}{
\llap{$^{\ANKARA}$}       METU-Middle East Technical University, TR-06531 Ankara, Turkey \\
\llap{$^{\ANNECY}$}       LAPP, Universit\'e de Savoie, CNRS/IN2P3, 74941 Annecy-le-Vieux, France\\
\llap{$^{\ASSERGILNGS}$}  Laboratori Nazionali del Gran Sasso dell'INFN, 67010 Assergi (L'Aquila), Italy \\
\llap{$^{\BARI}$}         Dipartimento di Fisica dell'Universit\`a  di Bari and INFN, 70126 Bari, Italy \\
\llap{$^{\BARIINFN}$}    INFN Sezione di Bari, 70126 Bari, Italy \\
\llap{$^{\BERN}$}         University of Bern, CH-3012 Bern, Switzerland \\
\llap{$^{\BOLOGNAINFN}$} INFN Sezione di Bologna, I-40127 Bologna, Italy \\
\llap{$^{\BOLOGNA}$}      Dipartimento di Fisica dell'Universit\`a  di Bologna and INFN, I-40127 Bologna, Italy \\\llap{$^{\BRUSSELS}$}     IIHE-Inter-University Institute for High Energies, Universit\'e Libre de Bruxelles, B-1050 Brussels, Belgium \\
\llap{$^{\DUBNA}$}        JINR-Joint Institute for Nuclear Research, 141980 Dubna, Russia \\
\llap{$^{\FRASCATI}$}     INFN - Laboratori Nazionali di Frascati, 00044 Frascati (Roma), Italy \\
\llap{$^{\FUNABASHI}$}    Toho University, 274-8510 Funabashi, Japan \\
\llap{$^{\HAMBURG}$}      Hamburg University, 22043 Hamburg, Germany\\
\llap{$^{\GAZWADONG}$}    Gyeongsang National University, 900 Gazwa-dong, Jinju 660-300, Korea\\
\llap{$^{\KARIYA}$}       Aichi University of Education, 448 Kariya (Aichi-Ken), Japan\\
\llap{$^{\KOBE}$}         Kobe University, 657 Kobe, Japan \\
\llap{$^{\LAQUILA}$}      Dipartimento di Fisica dell'Universit\`a  dell'Aquila and INFN, Gr. Coll. L'Aquila, Italy\\
\llap{$^{\LYON}$}         IPNL, Universit\'e Claude Bernard Lyon 1, CNRS/IN2P3, 69622 Villeurbanne, France\\
\llap{$^{\MOSCOWINR}$}    INR-Institute for Nuclear Research of the Russian Academy of Sciences, 117312 Moscow, Russia\\
\llap{$^{\MOSCOWITEP}$}   ITEP-Institute for Theoretical and Experimental Physics, 117259 Moscow, Russia \\
\llap{$^{\MOSCOWLPI}$}    LPI-Lebedev Physical Institute of the Russian Academy of Sciences, 117924 Moscow, Russia\\
\llap{$^{\MOSCOWSINP}$}   SINP MSU-Skobeltsyn Institute of Nuclear Physics of Moscow State University, 119992 Moscow, Russia \\
\llap{$^{\NAGOYA}$}       Nagoya University, 464-01 Nagoya, Japan\\
\llap{$^{\NAPOLIINFN}$}  INFN Sezione di Napoli, 80125 Napoli, Italy \\
\llap{$^{\NAPOLI}$}       Dipartimento di Fisica dell'Universit\`a Federico II di Napoli and INFN, 80125 Napoli, Italy \\
\llap{$^{\NEUCHATEL}$}    Universit\'e de Neuch\^atel, CH 2000 Neuch\^atel, Switzerland\\
\llap{$^{\OBNINSK}$}      Obninsk State University, Institute of Nuclear Power Engineering, 249020 Obninsk, Russia\\
\llap{$^{\PADOVAINFN}$}  INFN Sezione di Padova, 35131 Padova, Italy \\
\llap{$^{\PADOVA}$}       Dipartimento di Fisica dell'Universit\`a  di Padova and INFN, 35131 Padova, Italy \\
\llap{$^{\ROMA}$}         Dipartimento di Fisica dell'Universit\`a  di Roma ``La Sapienza" and INFN, 00185 Roma, Italy \\
\llap{$^{\SALERNO}$}      Dipartimento di Fisica dell'Universit\`a  di Salerno and INFN, 84084 Fisciano, Salerno, Italy \\
\llap{$^{\STRASBOURG}$}   IPHC, Universit\'e Louis Pasteur, CNRS/IN2P3, 67037 Strasbourg, France\\
\llap{$^{\URBINO}$}       CSAAE - Urbino University and INFN-Laboratori Nazionali di Frascati\\
\llap{$^{\UTSUNOMIYA}$}   Utsunomiya University, 320 Tochigi-Ken, Utsunomiya, Japan\\
\llap{$^{\ZAGREB}$}       IRB-Rudjer Boskovic Institute, 10002 Zagreb, Croatia\\
\llap{$^{\ZURICH}$}       ETH-Eidgen\"ossische Technische Hochschulen Z\"urich, CH-8092 Zurich, Switzerland \\
\llap{$^{a}$} Now at Chonnam National University\\
\llap{$^{b}$} Also PD, PINSTECH, P.O. Nilore, and COMSATS-CIIT, No. 30, H-8/1, Islamabad, Pakistan 
\\
}

\newcommand{\OperaAuthorList}{
A.~Anokhina$^{\MOSCOWSINP}$,
S.~Aoki$^{\KOBE}$,
A.~Ariga$^{\NAGOYA}$,
L.~Arrabito$^{\LYON}$,
D.~Autiero$^{\LYON}$,
A.~Badertscher$^{\ZURICH}$,
F.~Bay$^{\ANKARA}$,
A.~Bergnoli$^{\PADOVAINFN}$,
F.~Bersani~Greggio$^{\URBINO}$,
M.~Besnier$^{\ANNECY}$
% ...
\\}

\title{Study of the effects induced by lead on the emulsion films of the OPERA experiment}
\author{ \OperaAuthorList \\
         \OperaInstitutes \\
         E-mail: \email{patrizii@bo.infn.it}
         }

\abstract{The OPERA neutrino oscillation experiment is based on the use of the Emulsion Cloud Chamber (ECC).  In the OPERA ECC, nuclear  emulsion films acting as very high precision tracking detectors are interleaved with lead plates providing a massive target for neutrino interactions. We report on studies related to the effects occurring from the contact between emulsion and lead. A low radioactivity lead is required in order to minimize the number of background tracks in emulsions and to achieve the required performance in the reconstruction of neutrino events. It was observed that adding other chemical elements to the lead, in order to improve the mechanical properties, may  significantly increase the level of radioactivity  on the emulsions. A detailed study was made in order to choose a lead alloy with good mechanical properties and an appropriate packing technique so as to have a low enough effective radioactivity.}

\keywords{Nuclear emulsions; OPERA experiment; lead alloy;  polonium migration; $\alpha$-radioactivity}

\begin{document}

\section{Introduction}

\normalsize The OPERA experiment \cite{1} was mainly designed for the direct search of $\nu_{\tau}$ appearance in the pure $\nu_{\mu}$  CNGS beam from CERN to Gran Sasso \cite{2}. Its observation will give a definitive proof of the neutrino oscillation interpretation of the $\nu_{\mu}$ disappearance observed  first with atmospheric neutrinos [3-7]  and then in long baseline neutrino beams \cite{8}. 
The experiment is based on the use of  Emulsion Cloud Chambers (ECCs) \cite{kaplon} and of electronic detectors. In an ECC the nuclear emulsion  films act as very high precision tracking detectors, and are interleaved with plates of passive material. In the OPERA ECC the passive target material is lead. 
\par The OPERA detector is made of two identical supermodules. Each of them consists of a neutrino target section made of lead/emulsion ECC \textit{bricks},  and of a scintillator tracker detector, followed by a muon spectrometer \cite{1}. The total neutrino target mass  amounts to 1.35 kton. \par 
The basic unit of the target, the brick, is made of 57 emulsion films interleaved with 56 lead foils of 1 mm thickness. The emulsion films are made of 2 emulsion layers, each 43 $\mu$m thick, deposited on both sides of a 205 $\mu$m thick plastic base. The beam direction is perpendicular to the emulsion films. The brick has transverse dimensions of 12.8 x 10.3 cm$^{2}$,  a thickness of 8.1 cm and it weighs 8.3 kg. Fast automated microscopes are used for emulsion scanning \cite{9, 10}. \par

The nuclear emulsions consist of silver halide micro-crystals (each of them a few$\,$tenths of $\mu$m in size) dispersed in gelatin. The silver halide crystals are in the form of silver bromide crystals with a small amount of iodine in the crystal lattice. The density of OPERA emulsions is $\rho$ = 2.71 g/cm$^{3}$; their weight composition is given in Table \ref{tab:1}. \par

 \begin{table}
\begin{center}
{\small
\begin{tabular}
{|c|c|c|c|c|c|c|c|c|c|c|c|c|}\hline
 & {\bf Ag} & {\bf Br} & {\bf I} & {\bf C} & {\bf N} & {\bf O} & {\bf H} & {\bf S} & {\bf Si} & {\bf Na} & {\bf Sr} & {\bf Ba}  \\ \hline
{\bf A}  & 107.9 & 79.9 & 126.9 & 12 & 14 & 16 & 1 & 32 & 28 & 23 & 87.6 & 137.3 \\ \hline
{\bf Z} & 47 & 35 & 53 & 6 & 7 & 8 & 1 & 16 & 14 & 11 & 38 & 56 \\ \hline
{\bf \%} & 38.34 & 27.86 & 0.81 & 13.0 & 4.81 & 12.43 & 2.40 & 0.09 & 0.08 & 0.08 & 0.02 & 0.01  \\ \hline 
\end{tabular}
}
\end {center}
\caption {Composition of the OPERA emulsions (in \% by weight). K, F, Au, Fe and Cl nuclei are also present in very small amounts.} 
\label{tab:1}
\end{table}

Lead is well suited as passive material of the OPERA ECC bricks due to its high density and short radiation length: the former enhances the neutrino interaction rate and the latter the momentum determination by multiple Coulomb scattering as well as  electron identification and energy measurement by observing the shower development.
\par
\begin{figure}
\centering\includegraphics[width=.6\textwidth]{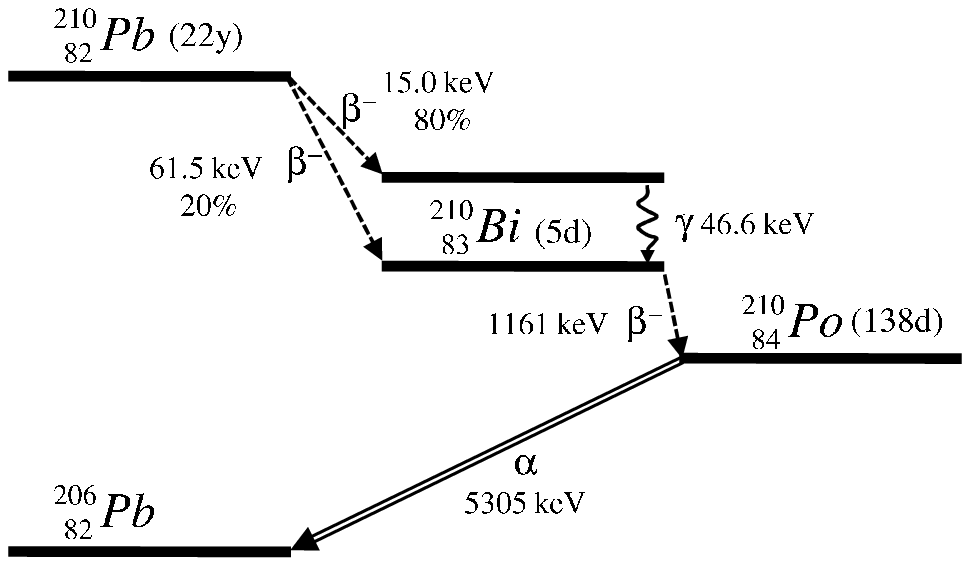}
\caption{The decay chain $^{210}Pb\rightarrow ^{210}Bi\rightarrow  ^{210}Po \rightarrow ^{206}Pb$. Within parentheses the half lives of the radioactive nuclides are given.}
\label{fig:decay-chain}
\end{figure}

\begin{figure}[!ht]
 \centering\includegraphics[width=.6\textwidth]{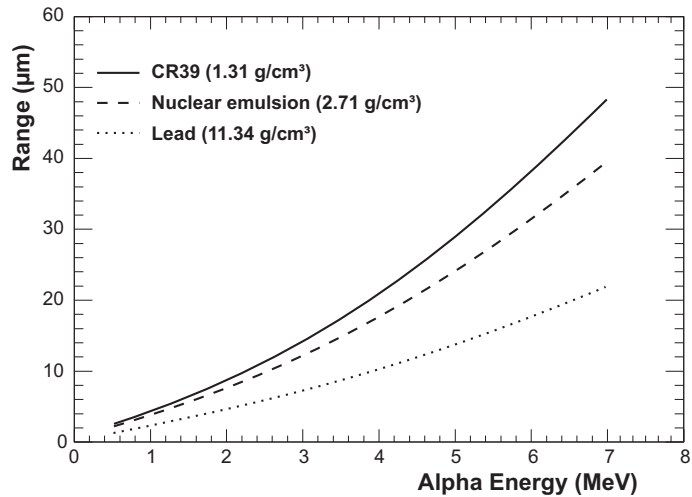}
 \caption{Ranges of alpha particles in nuclear emulsions, lead and CR39 as functions of the  particle energy.} 
\label{fig:ranges}
 \end{figure}

  Natural lead contains a contamination of the radioactive isotope $^{210}$Pb. Fig.$\,1$ shows the main decay chain from $^{210}$Pb $\rightarrow$ $^{206}$Pb.
Both alpha and beta decays produce background in nuclear emulsions and may affect the event reconstruction \cite{11}. Alpha-particles of 5.3 MeV are emitted from the decay of $^{210}$Po. Fig.$\,$\ref{fig:ranges} shows the  ranges of $\alpha$ particles in nuclear emulsions, lead, and CR39\footnote{CR39 is a polymer nuclear track detector (see section 2.4)} as functions of the particle energy. The ranges  were computed using the SRIM-2003 Monte Carlo code \cite{12}. The 5.3 MeV $\alpha$'s from $^{210}$Po decay have a range of 25 $\mu$m in nuclear emulsion. The $\beta$-radioactivity arises from the decay chain $Pb^{210}\stackrel{\beta(61.5, 15.0)}{\longrightarrow}Bi^{210}\stackrel{\beta(1161)}{\longrightarrow}Po^{210}$
(see Fig.\ \ref{fig:decay-chain}). Low and high energy $\beta$s come from the decay of $^{210}$Pb and of $^{210}$Bi, respectively. \par

If $\alpha$-emitters are uniformly distributed inside the lead plate, because of the energy loss in lead the $\alpha$-range and energy distributions in the emulsions should be  almost flat. However,
in pure lead and in some lead alloys $^{210}$Po migrates to the surface layers of the lead plate \cite{13} so that all $\alpha$-particles emitted towards the emulsion have a high probability of being detected. The alpha-activity  is therefore enhanced. \par

We made an extensive search in order to select the least radioactive lead on the market. The DoeRun brand had a bulk activity of $\sim$27$\,$Bq/kg, but it was not available in sufficient quantity. The lead finally used in OPERA was the Britannia brand, having a bulk activity of $\sim$$\,$80$\,$Bq/kg. \par

Different types of lead alloys
were obtained by adding small percentages of Ca or Sb to achieve mechanical rigidity and were tested for the OPERA bricks. 
 
We evaluated the compatibility in relation to the packing technique of the lead and the emulsions, by measuring the  single isolated grains (\textit{fog}) after the emulsion development. The effect depends on the lead alloy. \par

A detailed study was made in order to choose a lead alloy with good mechanical properties and an appropriate packing technique so as to have low enough effective radioactivity and acceptable fog on the emulsions.
We eventually decided to use the PbCa 0.03$\%$ alloy which presents good mechanical properties and was found to be safe from the point of view of  $\alpha$-radioactivity and  compatibility with nuclear emulsions.\par

In Section 2 the measurements of the energy, ranges and rates of $\alpha$-particles emitted from lead are presented, together with a brief description of the different methods used; the measurements are compared to Monte Carlo calculations. The  effects related to the brick packaging are discussed in Section 3. The effects of the lead $\alpha$- and $\beta$-activity on the OPERA emulsion scanning and reconstruction efficiency are studied in Section 4. The conclusions are given in Section 5.

\section{Measurements of $\alpha$-radioactivities }

The $\alpha$-radioactivity of the lead samples to be tested was studied using different techniques: 

(i)   nuclear emulsions; 

(ii)  surface barrier silicon detectors (SBSiD); 

(iii) 2$\pi$ gas flow proportional counter;  

(iv) CR39 nuclear track detectors. 

\subsection{Measurements with nuclear emulsions}

In nuclear emulsions the passage of an ionizing particle through a micro-crystal renders it developable by chemical agents. After development, followed by fixing and washing to remove undeveloped crystals, the gelatin is transparent and the paths of charged particles are visible as trails of small dark silver grains, with a diameter smaller than 1 $\mu$m.  A track in an emulsion layer of an emulsion film is a sequence of aligned grains and is called \textit{microtrack}.  A \textit{base track} is a segment which connects  two microtrack ends across the plastic base of the same film. It is characterised by a higher angular resolution because it is less affected by  distortions and because of the larger lever arm. A \textit{ volume track} connects at least three base tracks observed in different emulsion films. \par
	
Alpha particles of 5 MeV produce in the emulsions thick (\textit{black}) tracks, which are not easily reconstructed as tracks by automated microscopes. They were measured semi-automatically using an objective mounted on the automated microscopes used in OPERA \cite{9, 10}. Track lengths were determined by connecting the first and last detected grains. The emulsions were developed after several weeks in order to monitor the lead radioactivity and its time dependence. In order to compare different lead samples (pure Pb, PbSb and PbCa alloys), sandwiches of lead and emulsion were prepared. Co-laminated lead samples (consisting of 1 mm thick lead alloy plates with a 30 $\mu$m pure DoeRun lead layer co-laminated on both sides) were also  tested in order to investigate a possible reduction of the  $\alpha$-emission effect. \par
The range distribution of $\alpha$-particles in nuclear emulsions exposed for 52 days to the $\alpha$-radioactivity of a lead PbSb 2.5$\%$ sample is shown in Fig.$\,$3. It is compatible with a surface activity due to $^{210}$Po migration, as almost all the $\alpha$`s have the 25 $\mu$m range corresponding to the 5.3 MeV energy expected from $^{210}$Po decay in absence of energy loss in lead. Notice that ranges as little as a few$\,$$\mu$m were measured.  For PbCa alloy samples no peak was observed but only few events for ranges smaller than 10 $\mu$m.\par

\begin{figure}[!ht]
 \centering \includegraphics[width=.6\textwidth]{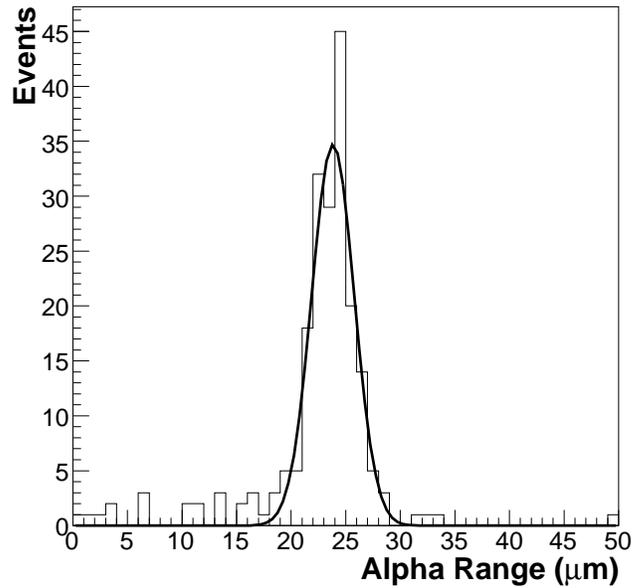}
 \caption{Ranges of $\alpha$-particles measured semi-automatically in nuclear emulsion films which were in contact with a PbSb 2.5$\%$ sample for 52 days. The peak at a range of 25 $\mu$m is compatible with what is expected from a surface emission of 5.3 MeV $\alpha$`s from $^{210}$Po decay. } 
\label{fig:measured-range}
 \end{figure}

Table 2 gives the surface activity (tracks cm$^{-2}$ d$^{-1}$) measured for different  samples and different times after lead production. The quoted errors are statistical. Systematic uncertainties are of the order of 10$\%$.  \par

\begin{table}
\vspace{0.4cm}
\begin{center}
\begin{tabular}{|c|c|c|c|c|c|c|}
\hline
Primary & Secondary & Colamination & Sample & \multicolumn{3}{|c|}{Time after
production} \\ 
lead & component & lead & thickness & \multicolumn{3}{|c|}{(weeks)} \\
 & (\%) & & (mm) & \multicolumn{3}{|c|}{ } \\
\hline
\multicolumn{4}{|c|}{ } & 4 & 4 & 16 \\
\cline{5-7}
\multicolumn{4}{|c|}{ } & \multicolumn{3}{|c|}{Surface activity (tracks cm$^{-2}$ d$^{-1}$)} \\
\hline
DoeRun & None & None & 3 & $19.0 \pm 0.5$ & $22 \pm 3$ & $31 \pm 2$ \\
\hline
Britannia & None & None & 3 & $38 \pm 1$ & $32 \pm 3$ & $31 \pm 2$ \\
\hline
Britannia & Ca (0.07) & None & 1 & $7.0 \pm 0.5$ & $7 \pm 3$ & $14 \pm 1$ \\
\hline
Britannia & Ca (0.03) & None & 1 & - & $10.0 \pm 0.5$ & $25 \pm 1$ \\
\hline
Britannia & Ca (0.03) & DoeRun & 1 & - & $8.0 \pm 0.5$ & $14 \pm 1$ \\
\hline
Britannia & Sb (2.5) & None & 1 & $65 \pm 2$ & $74 \pm 3$ & $131 \pm 3$ \\
\hline
Britannia & Sb (2.5) & DoeRun & 1 & $71 \pm 3$ & $82 \pm 3$ & $112 \pm 3$ \\
\hline
\end{tabular}
\vspace{0.4cm}
\caption{$\alpha$-activity  for different types of Pb and Pb alloys at different times after lead production, measured using nuclear emulsions.}
\label{tab:2}
\end{center}
\end{table}

From the above measurements one can conclude that:
 
(i) the large $\alpha$-activity observed in PbSb 2.5$\%$ comes from a surface activity; 

(ii) the $\alpha$-activities for PbCa 0.07$\%$ and 0.03$\%$ are much smaller than in PbSb 2.5\% and do not come from a surface activity; 

(iii) for pure lead samples the DoeRun lead is, right after production, less radioactive than the Britannia lead but the difference tends to decrease with time indicating a migration effect; 

(iv) for PbCa 0.07\% or 0.03\% alloys, the activity is lower than for pure lead and the $^{210}$Po migration seems to be limited. \par
(v) co-lamination of PbSb alloy samples with pure Pb does not bring any advantage in terms of reduction of the surface $\alpha$-activity. The measurements indicated the presence of large spatial fluctuations, which could be due to a non perfect co-lamination; furthermore a good uniform co-lamination is technically difficult to achieve.

\subsection{Measurements of $\alpha$-activity with a 2$\pi$ gas flow proportional counter}

The $\alpha$-radioactivity of different samples of pure lead and of lead alloys was measured using a 2$\pi$ gas flow proportional counter of 4.9 cm$^{2}$ sensitive area. The results of the measurements are summarized in Table 3. They confirm the lower activities of PbCa samples with respect to the PbSb ones, both for co-laminated and non co-laminated foils.

\begin{table}
\begin{center}
\begin{tabular}{|c|c|c|c|c|c|c|c|c|}
\hline
Primary & Secondary & Colamination & Sample & \multicolumn{5}{|c|}{Time after
production} \\ 
lead & component & lead & thickness & \multicolumn{5}{|c|}{(weeks)} \\
 & (\%) & & (mm) & \multicolumn{5}{|c|}{ } \\
\hline
\multicolumn{4}{|c|}{ } & ~~~0~~~ & ~~~2~~~ & ~~~3~~~ & ~~~9~~~ & 17 \\
\cline{5-9}
\multicolumn{4}{|c|}{ } & \multicolumn{5}{|c|}{Surface activity (counts cm$^{-2}$ d$^{-1}$)} \\
\multicolumn{4}{|c|}{ } & \multicolumn{5}{|c|}{($10 - 20 \%$ error)} \\
\hline
Britannia & Sb (2.5) & None & 1 & 65 & 84 & 79 & 82 & \\
\hline
Britannia & Sb (2.5) & DoeRun & 1 & 28 & 44 & 35 & 52 & \\
\hline
Britannia & Ca (0.06) & None & 1 & - & - & - & 12 & 12 \\
\hline
Britannia & Ca (0.06) & DoeRun & 1 & - & - & - & 14 & 10 \\
\hline
DoeRun & None & DoeRun & 1 & 21 & - & - & - & - \\
\hline
Britannia & None & None & 3 & 33 & - & - & - & - \\
\hline
Britannia & Ca (0.07) & None & 3 & 6 & - & - &  & - \\
\hline
Britannia & Sb (2.5) & None & 3 & 42 & - & - & - & - \\
\hline
\end{tabular}
\vspace{0.6cm}
\caption{$\alpha$-activity  for different types of Pb and Pb alloys and times after lead production, measured using a 2$\pi$ gas flow proportional counter.}
\label{tab:3}
\end{center}
\end{table}

\subsection{Measurements with Surface Barrier Silicon Detectors }
Surface Barrier Si Detectors (SBSiDs), so called because the active volume (\textit{depleted region}) is very close to the surface of the detector, allow $\alpha$-particle spectroscopy to the level of 30 keV resolution with about 100$\%$ efficiency. Two different commercial Surface Barrier Silicon detectors (SBSiD-1 and SBSiD-2) were used in two different OPERA labs. The distance of the lead sample from the detector was $\sim$2 mm and  $\sim$10 mm for SBSiD-1 and -2, respectively. The measurements were done inside a vacuum chamber ($\sim$$10^{-5}$ mbar). The  energy was calibrated using  mixed $\alpha$-sources of $^{273}$Np, $^{241}$Am and $^{244}$Cm for the SBSiD-1, and of $^{239}$Pu, $^{241}$Am and $^{244}$Cm for the SBSiD-2. The calibration spectra are shown in Fig.$\,$4.

Fig.$\,$5 shows the energy spectra obtained for six different lead samples. Notice that for the PbSb 2.5\% samples the events cluster around an energy of 5.3 MeV, confirming the nuclear emulsion data. The noise level is very low for energies above 1.5 MeV; below 1.5 MeV there is some general noise which increases with decreasing energy. In Fig.$\,$5 one observes only a minor structure at E $\sim$$\,$5.3 MeV 
for PbCa 0.03\%; no structure is observed for PbCa 0.07\%. From these measurements one can confirm the preliminary conclusions reached at the end of Section 2.1, in particular that the $\alpha$-activity from $^{210}$Pb in PbCa 0.07\% and 0.03\% is much smaller than in PbSb 2.5\%. 

\begin{figure}[!ht]
 \centering\includegraphics[width=1.\textwidth]{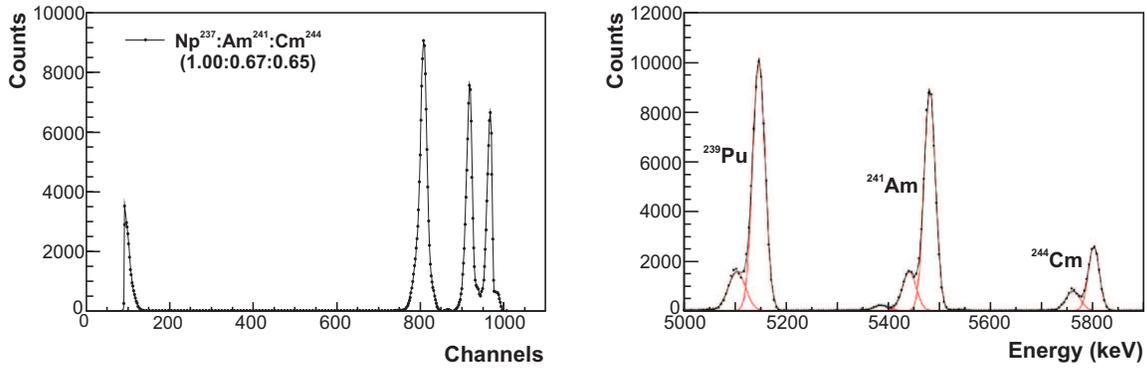}
\caption{Left: calibration of SBSiD-1  with a mixed $\alpha$-source of  $^{273}$Np (4.79 MeV), $^{241}$Am (5.49 MeV) and  $^{244}$Cm (5.81 MeV). Right: calibration of SBSiD-2 with a mixed $^{239}$Pu (5.16 MeV), $^{241}$Am (5.49 MeV) and $^{244}$Cm (5.81 MeV) alpha source. } 
\label{fig:SBDcalib}
\end{figure}

\begin{figure}[!ht]
 \centering\includegraphics[width=.9\textwidth]{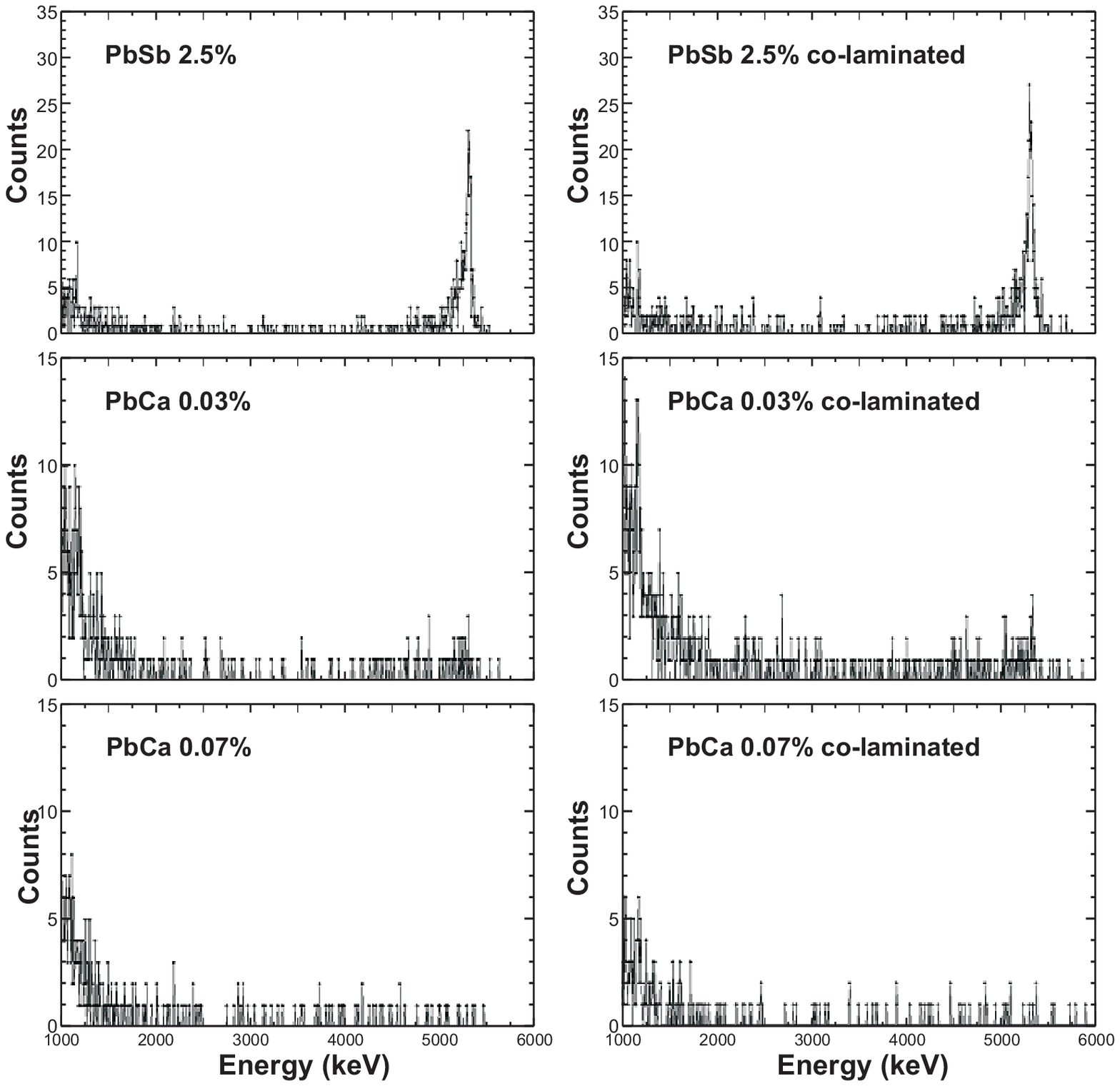}
\caption{ $\alpha$-radioactivity measurements with a SBSiD for six lead samples: on the left panels, from top  PbSb 2.5$\%$,  PbCa 0.03$\%$, and  PbCa 0.07$\%$ samples; on the right panels, from top, co-laminated PbSb 2.5$\%$,  PbCa 0.03$\%$, and  PbCa 0.07$\%$ samples.} 
\label{fig:SBDspectra}
 \end{figure}

Some PbSb 2.5\% samples were chemically etched so that a thickness of 3 to 90 $\mu$m of surface material was removed. The $\alpha$-radioactivity was measured before and at different times after etching. Some results are shown in Fig.$\,$6. The surface radioactivity  decreases immediately after etching but then it increases with time. For example, the $\alpha$-radioactivity was 18 tracks cm$^{-2}$ d$^{-1}$ right after the removal of 27 $\mu$m of total surface thickness, 28 tracks cm$^{-2}$ d$^{-1}$ 7.3 weeks after etching and 41 tracks cm$^{-2}$ d$^{-1}$  14.5 weeks after etching. This indicates an ongoing diffusion of the $^{210}$Po towards the lead surface.\par

Surface Barrier Silicon Detectors measurements were used also to evaluate the time variation of $\alpha$-radioactivity; the results  are shown in Fig.$\,$7 with those of CR39 detectors and discussed in  Section 2.5. \par

\begin{figure}[!ht]
 \centering\includegraphics[width=.75\textwidth]{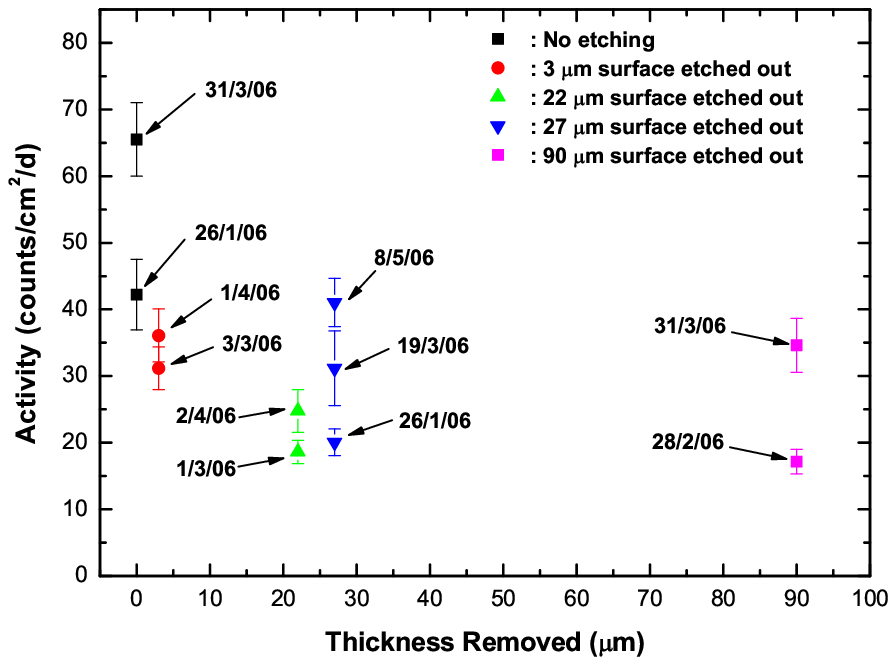}
 %\vspace{-1.cm}
\caption{Surface $\alpha$-activity for PbSb 2.5$\%$ lead samples  versus the surface thickness removed by etching. The measurements were made at different times after etching, using a surface barrier silicon detector. } 
\label{fig:etch-samples}

 \end{figure}

\subsection{Measurements with CR39 nuclear track detectors}
 In passive nuclear track detectors (NTDs) the damage produced by a charged particle can be enlarged through chemical etching and thus made visible under an optical microscope as a conical \textit{etch-pit} (called a \textit{ track}). The etch-pit size is related to the particle Restricted Energy Loss (REL), i.e. the energy released inside a cylindrical region of about 10 nm diameter around the particle trajectory  \cite{14,15}.
The polymer  CR39 is a widely used NTD. CR39  sheets\footnote{ The CR39 was manufactured by the Intercast Europe, Spa, Parma, Italy. CR39 is a trademark of PPG industries. It is a polymer widely used for making lenses for sunglasses. The Intercast CR39 used here as NTD was made in a specific scientific production line. } were first  \textit{refreshed} by etching them in a solution 8N KOH + 3$\%$ alcohol at 75$\,$$^{\circ}$C in order to remove previously recorded tracks from environmental radioactivity, mainly $\alpha$-tracks from radon \cite{16}. About 120 $\mu$m were removed from each side of the CR39 sheets; this thickness is large with respect to the 33.5 $\mu$m range of $\alpha$-particles from radon in CR39. \par 
 The refreshed CR39 sheets were then placed in contact with the Pb samples. After exposure, the CR39 sheets were etched in a solution 6N NaOH at 70$\,$$^{\circ}$C for 6 hours. The CR39 sheets were scanned and $\alpha$-tracks counted by a semi-automated optical microscope at a  magnification of   250$\times$. A global efficiency of $\sim$ 65\% for the detection of 5.3 MeV  $\alpha$'s was estimated; the detector energy threshold for  
$\alpha$-particles was about 3 MeV. \par
 The results of the measurements are given in Fig.$\,$7 and discussed in the following Section. \par

\subsection{Time evolution of $\alpha$-radioactivity }
The time dependence of the $\alpha$-activities of PbSb 2.5$\%$, PbCa 0.07$\%$ and PbCa 0.03\% measured over $\sim$1 year using SBSiDs and CR39 NTDs are plotted in Fig.$\,$7 versus the time elapsed since the production of the lead alloy plates. The errors are statistical standard deviations. The solid lines are fits of the data to the expression $y = y_{0} + y_{1}e^{-t/\tau}$ ($\tau \sim$$\,$few months). For SBSiDs only the counts in  the $4.0 < E_{\alpha} < 5.5$ MeV interval are considered. The activity in the first 70 days increases by about 40$\%$ for the PbSb 2.5\% alloy; it increases by 10-20$\%$ for PbCa 0.07\% and 0.03\% alloys over a longer time interval. Over longer times the radioactivity of PbCa 0.07\% and 0.03\% increased a little further remaining at a very low level. The $\alpha$-radioactivity in PbSb 2.5\% increased further and then remained constant at values $\sim$$\,$25$\,$times larger than for PbCa 0.07\% and  0.03\%. Fig.$\,$7b shows that the $\alpha$-activity in PbSb 2.5\% co-laminated with 30 $\mu$m of pure DoeRun Pb behaves approximately as for  PbSb 2.5\% without co-lamination; therefore co-lamination does not reduce the surface $\alpha$-activity. 
\par These results give quantitative information on the time dependence of the $\alpha$-radioactivity and confirm the preliminary conclusions reached using nuclear emulsions, presented in Section$\,$2.1. \par

\subsection{Overview on the $\alpha$-radioactivity measurements}

The samples with the lowest $\alpha$-radioactivity are those with PbCa$\,$0.07\% and 0.03\%. For these samples there is apparently no enhanced surface radioactivity. A possible explanation is that in the PbSb 2.5\% samples the $^{210}$Po atoms lie in a global potential which has a minimum at the edges of the sample. For the PbCa 0.07\% and 0.03\% samples it may be that the production process reduced the radioactivity by evaporation of $^{210}$Po  and/or that the global potential is more uniform. \par
 
\begin{figure}
\begin{center}
\resizebox{!}{8.cm}{\includegraphics{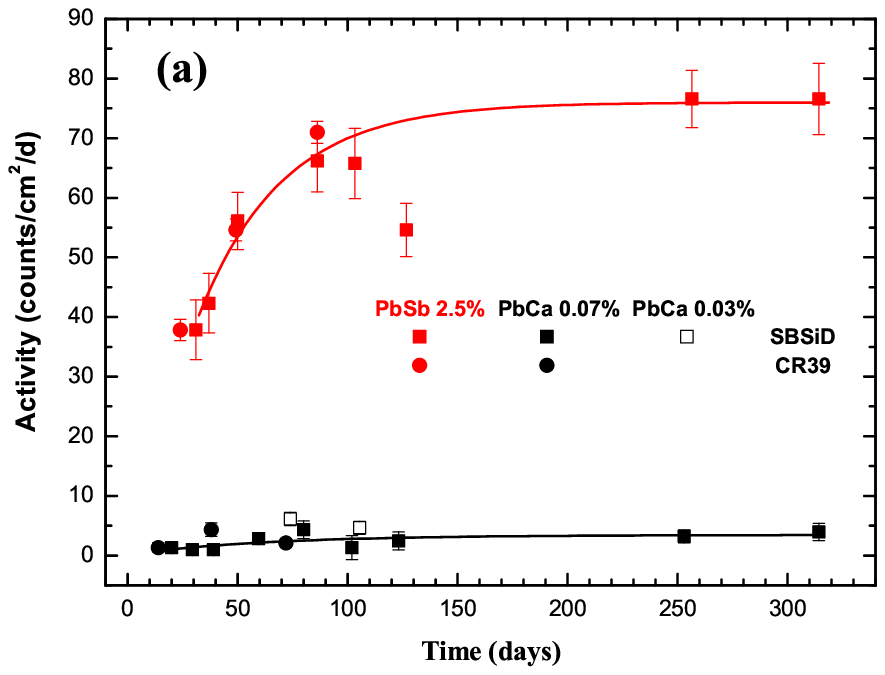}}
%\hspace{0.2cm}
\resizebox{!}{8.cm}{\includegraphics{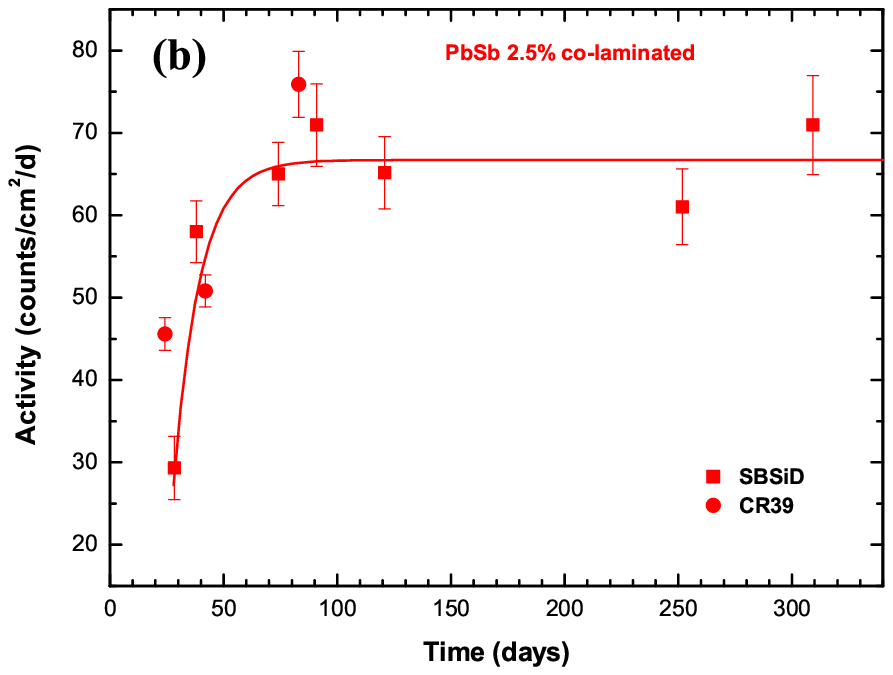}}
\caption{Time dependence of the alpha-activity from $^{210}$Po (a) in PbSb 2.5$\%$, PbCa 0.07$\%$, PbCa 0.03$\%$ and (b) in  PbSb 2.5$\%$ co-laminated with 30 $\mu$m of pure DoeRun lead. The solid lines are fits to the data (see text). }
\label{fig:activity-vs-time}
\end{center}
\end{figure}

\subsection{Monte Carlo simulations}
In order to better understand the features of the OPERA lead radioactivity,  Monte Carlo (MC) simulations of the $^{210}$Po $\beta$-decay and of the $^{210}$Po  5.3 MeV $\alpha$-decay inside the OPERA lead plates were made. Two set-ups were considered:\par
-   a single lead plate with an area of 10x10 cm$^{2}$ in contact with an emulsion film with the same surface ($\alpha$-study); \par
-   an OPERA brick, to evaluate the electron rate from $\beta$-decays.\par

The simulations were done using the GEANT 3.2 package.
The emulsion radioactivity rate is defined by counting the arrival of an  $\alpha$-particle inside the sensitive volume of the first emulsion layer. This implies a low threshold corresponding to an alpha range of 1 $\mu$m. We assumed for lead an activity of 80$\,$Bq/kg (Britannia lead). \par
 
If the $\alpha$s are generated uniformly in the lead one obtains a \textit{flat} $\alpha$ range distribution in emulsion as shown in Fig.$\,$8a. Figs.$\,$8b and 8c show the $\alpha$ emission points inside the lead and the energy spectrum of $\alpha$s entering the emulsion, respectively.  
If the $\alpha$-emission is from a thin surface layer the $\alpha$-range distribution in emulsion is as shown in Fig.$\,$9 for different thicknesses of the  emitting layer. A comparison of the simulations shown in Fig.$\,$9 with the measured spectra for PbSb 2.5\% (Fig.$\,$3) indicates that $\alpha$s are emitted from a surface thickness of $<$1 $\mu$m of the PbSb 2.5\% alloy. \par
A possible solution to protect the emulsion from the $\alpha$s would be to interpose foils of at least 12 $\mu$m of steel or 17 $\mu$m of low radioactivity lead, which would stop the 5.3 MeV $\alpha$s. This was the original purpose of the lead co-lamination. \par

\begin{figure}[!ht]
 \centering\includegraphics[width=.8\textwidth]{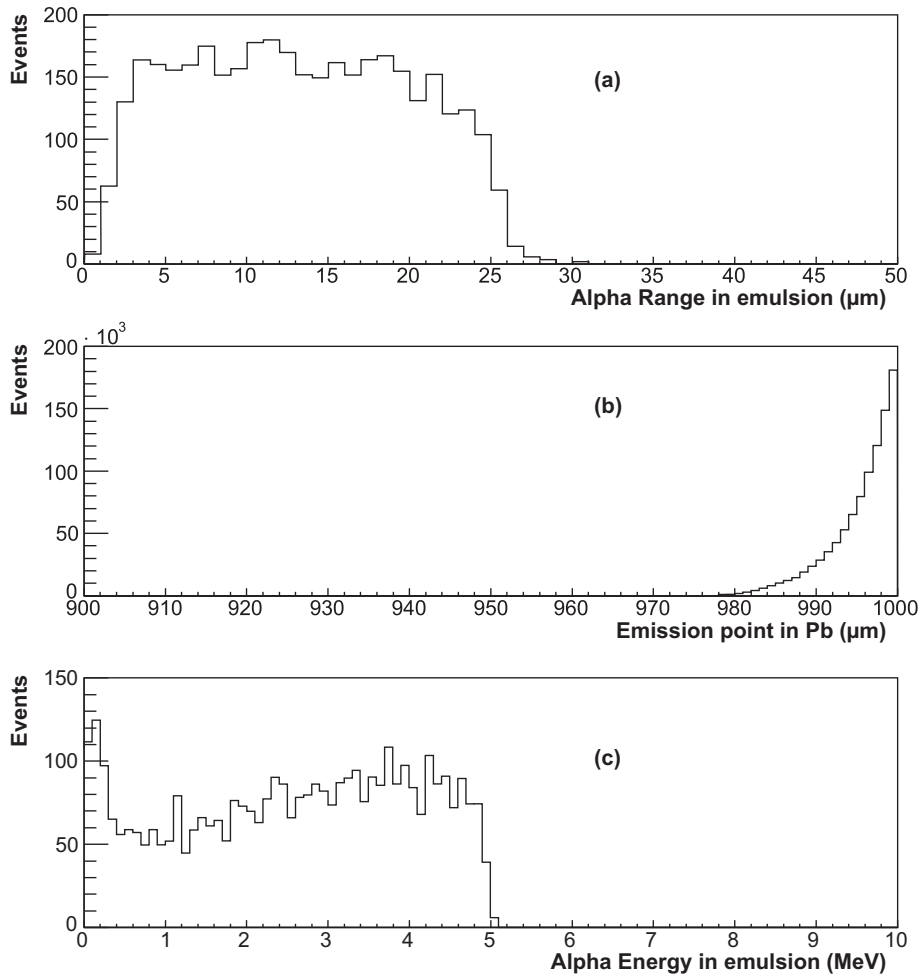}

 \caption{(a) Monte Carlo predicted range of alphas in nuclear emulsions for a uniform distribution of the $\alpha$-activity in the lead: the distribution is flat, very different from the measured spectra from PbSb 2.5\%, which indicate a surface effect. (b) Monte Carlo $\alpha$ emission point in lead: only alphas emitted from the last 20 $\mu$m of lead are detected in emulsions. (c) Monte Carlo energy spectrum of the alphas entering the emulsion, reflecting the slowing down in the last 10 $\mu$m of lead. } 
\label{fig:MC1}
 \end{figure}

\begin{figure}[!ht]
 \centering\includegraphics[width=.8\textwidth]{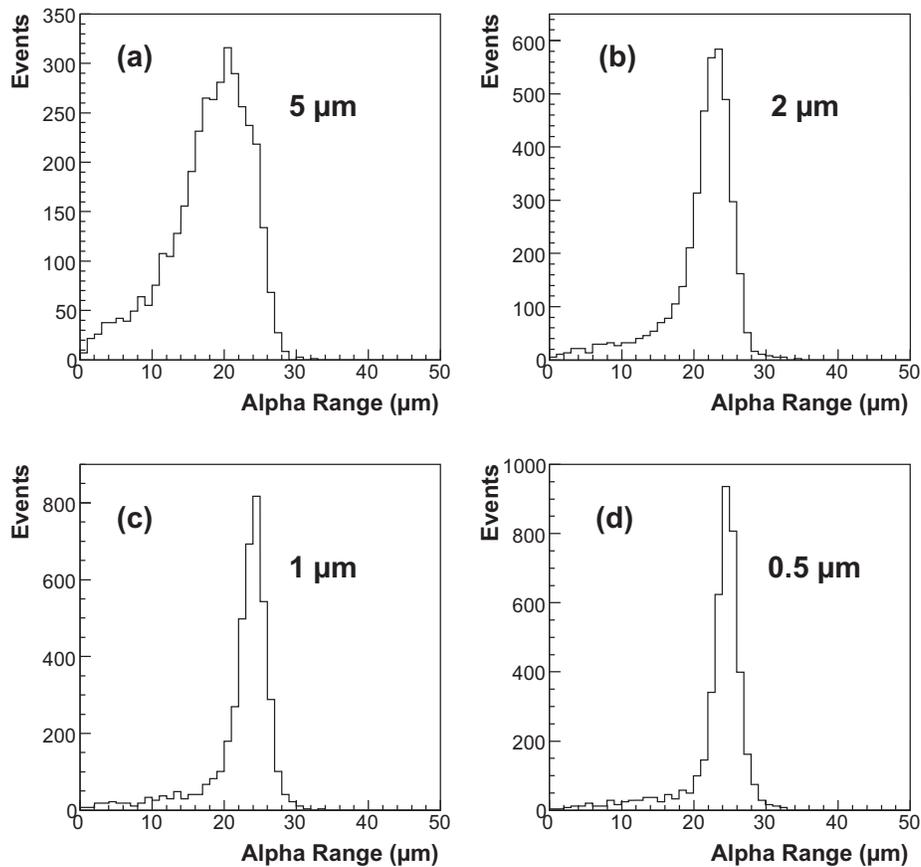}
 \caption{ Monte Carlo studies: the range in emulsion of alphas emitted from different lead surface radioactive thicknesses: (a) 5 $\mu$m, (b) 2 $\mu$m, (c) 1 $\mu$m and (d) 0.5 $\mu$m.  } 
\label{fig:MC2}
 \end{figure}

\begin{figure}[!ht]
\hspace {-0.5cm}\centering\includegraphics[width=1.\textwidth]{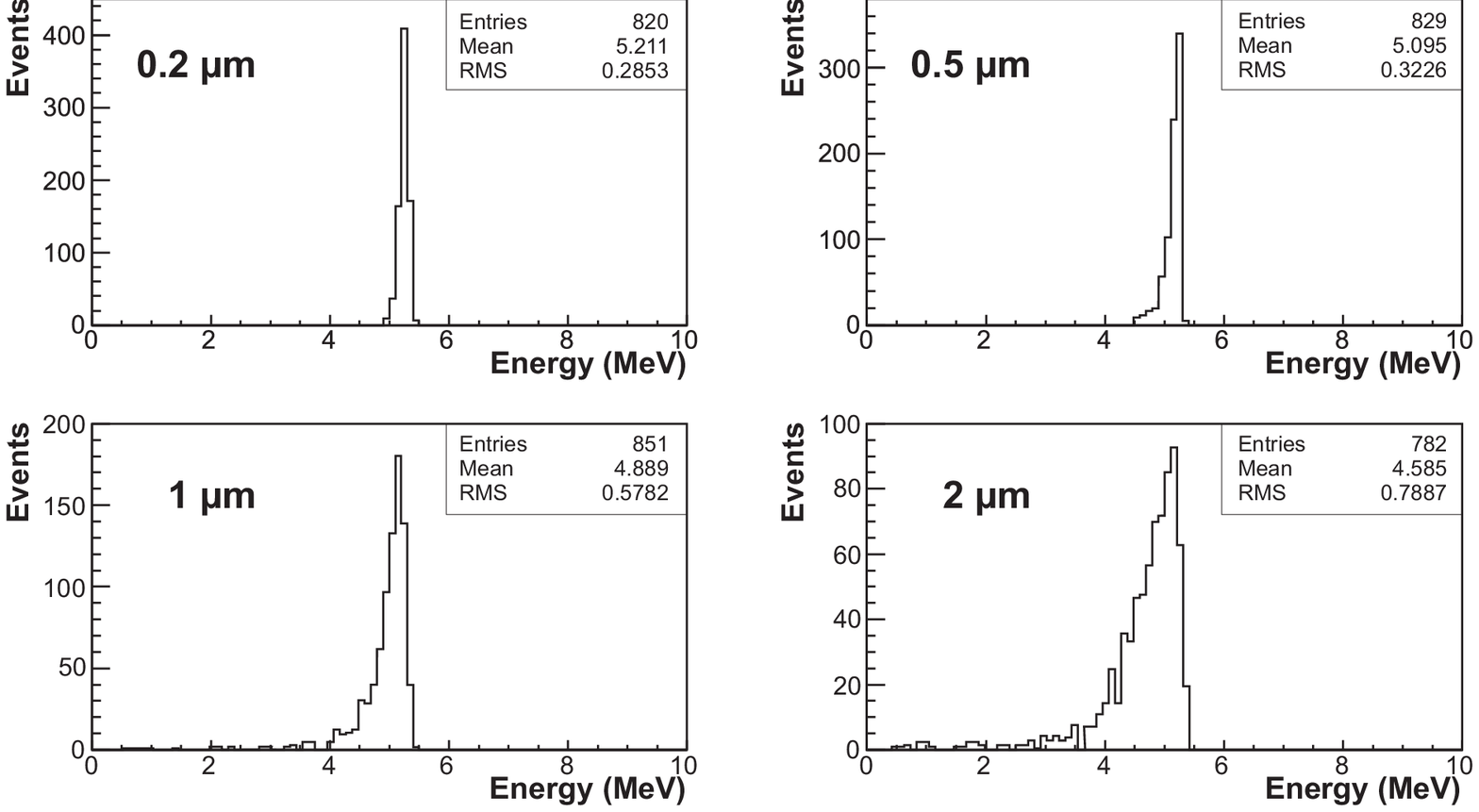}

 \caption{ Monte Carlo predicted energy spectra in silicon surface barrier detectors, depending on the $^{210}$Po effective thickness of the surface emission region. } 
\label{fig:MC3}
 \end{figure}

The energy spectra obtained by SBSiDs were simulated by considering different thicknesses of the surface emission regions: 0.2 $\mu$m, 0.5 $\mu$m, 1 $\mu$m, and 2 $\mu$m. An experimental resolution of 50 keV was folded in the distributions shown in Fig.$\,$10.  A further comparison of these simulations with the experimental data (Fig.$\,$5) allows us to state that the surface emitting region in PbSb 2.5\% alloys has a thickness  $<$0.5 $\mu$m. \par

Monte-Carlo simulations of lead $\alpha$-radioactivity predicted rates of $\beta$-tracks in emulsion is 200 $\beta$-tracks cm$^{-2}$ d$^{-1}$ for the single lead plate in contact with an emulsion film, and 622 $\beta$-tracks cm$^{-2}$ d$^{-1}$ for emulsions inside a brick.

\section{Effects related to the brick packaging} 
Tests were performed to investigate possible effects induced by the contact of the emulsions with different types of lead plates. For this purpose the fog in the emulsion, defined as the number of isolated dark grains per unit volume (grains per 1000 $\mu$m$^{3}$), was measured in the emulsion film in contact with the lead plates, in contact with other emulsion films and in  emulsion films not in contact with lead. \par
In past tests an increase of the fog was observed in nuclear emulsions  packed in vacuum with PbCa lead alloy plates; on the other hand if the emulsion-lead contact was in air  no fog increase was observed [12]. \par

In order to further investigate the above effects, four OPERA-like bricks were assembled in different conditions using PbCa 0.07\% lead plates: two bricks were packed $\grave{a} \: la$ OPERA using the standard OPERA brick cover (\textit{spider}),  one brick was sealed with a very small amount of air (10 cc),  the other one was left open; the other two bricks were assembled without the OPERA brick spider, one open and one sealed with 10 cc of air. The bricks were all stored in an oven at 35$^{\circ}$C. The fog was  also measured in  the emulsion film packed under vacuum without any lead and stored in the same oven with the four bricks. \par
 After 5 weeks, the emulsion layer placed at the bottom side of the brick (\textit{bottom}), the middle one (\textit{middle}) and the one at the top (\textit{top}) were developed and the fog density was measured. The results are shown in Fig.$\,$11. The fog measurements performed automatically and checked manually gave values from 4 to 6.5 grains/1000 $\mu$m$^{3}$, which  are smaller than the fog in the emulsion packed under vacuum  without lead.

\begin{figure}[!ht]
 \centering\includegraphics[width=.7\textwidth]{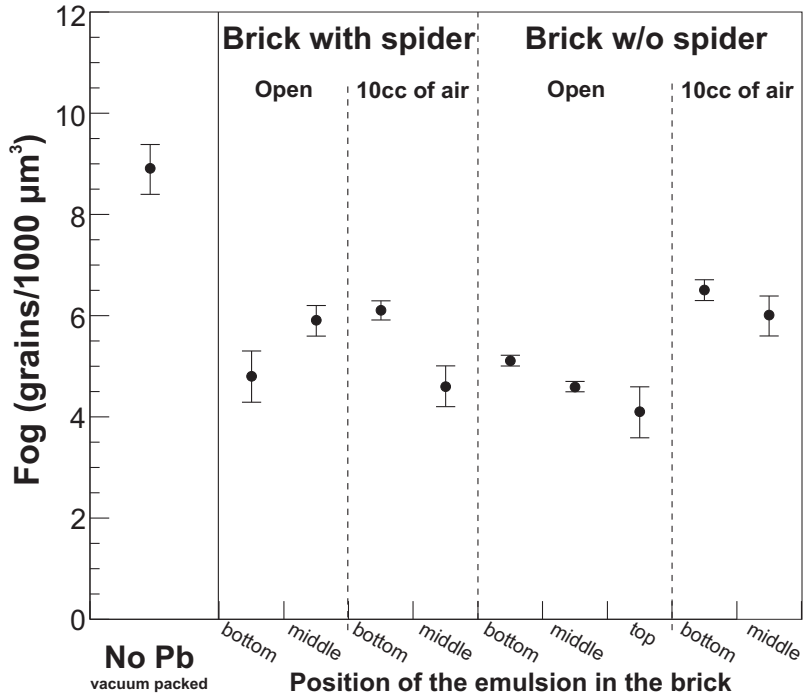}
\vspace{0.3cm}
 \caption{Fog measurements in emulsions placed in an OPERA-like brick in contact with PbCa 0.07\% plates and stored for 5 weeks in an oven at 35$^{\circ}$C. From the left: emulsion packed under vacuum, without any lead, brick packed using the OPERA brick  spider (open or with 10 cc of air),  brick packed without the spider  (open or with 10 cc of air). Error bars are total errors obtained by adding in quadrature systematic and statistical uncertainties (measurements performed automatically with an ESS).
} 
\label{fig:fog}
 \end{figure}

A \textit{half-brick }(23 emulsion films and 23 lead plates) was assembled using PbSb 2.5\% and maintained for 8 months in the Gran Sasso underground laboratory inside a shield of very pure copper and lead \cite{11}. One emulsion in contact with the lead plate was developed and analysed: its outer surface was full of $\alpha$-tracks and had a very low fog density of $\sim$$\,$2.9 grains/1000 $\mu$m$^{3}$. This low fog level was probably connected with the ambient conditions of the underground lab.

\section{ Effects of $\alpha$- and $\beta$-radioactivity on emulsion scanning. }
To assess the effect of $\alpha$ and $\beta$-radioactivity on emulsion scanning, several tests were performed by exposing emulsions to different $\alpha$- and $\beta$-radioactive sources. We also used 500 MeV electrons from the INFN-LNF electron beam and 1 GeV pions at the KEK accelerator. The $\alpha$- and $\beta$-exposures were performed to mimic the PbSb 2.5$\%$ effect from 2 to 5 years of contact. 
\par We evaluated the impact of different densities of $\alpha$- and $\beta$-tracks on the performance of the automated miscroscopes described in Ref.$ \, $[11] in terms of background and efficiency. The acquisition speed (20 cm$^{2}$/hours) did not decrease even with an $\alpha$ dose of 400 $\alpha$/view but the number of  fake  microtracks increased by more than a factor of  2.5 with respect to the unexposed zones. Almost all the reconstructed base tracks in the $\alpha$- and $\beta$-exposed zones were \textit{bad tracks} and most of them could be rejected with a simple quality cut;  the radiation exposure increased the base track density up to a factor of 2. The effects will be considerably smaller in the case of PbCa 0.07$\%$ or 0.03$\%$\par

Overall, we have not encountered problems on the emulsion scanning at the level of track finding. Further investigation will be carried out in order to evaluate the effects on the interaction vertex reconstruction and event analysis.

\section{Conclusions}

In the OPERA neutrino oscillation experiment, nuclear emulsion films are in contact with lead plates and packed together in a basic detector unit called \textit{brick}. The lead with the lowest radioactivity practically available on the market was chosen. In order to ensure adequate mechanical properties, other chemical elements (like Ca or Sb) are added to the lead. 

The effective radioactivity acting on the emulsion films was found to depend on the type and on the concentration of the added chemical element. In particular, evidence was found for the occurrence of a migration of $^{210}$Po towards the surface of PbSb 2.5\% samples, in contact with the emulsions. Monte Carlo calculations give results consistent with this interpretation of the high radioactivity which, in certain conditions, affects the emulsions. The comparison of experimental data with Monte Carlo simulations indicated that the thickness of the $\alpha$-emitting region in PbSb 2.5\% samples is smaller than 0.5 $\mu$m.\par

The main measurements are summarized as follows.
The $\alpha$-radioactivity is  lower for the PbCa 0.07\% and 0.03\% alloys. For PbSb 2.5\% it is 
$\sim$25$\,$times larger than in PbCa 0.07\%. An increase with time of the surface activity of PbSb 2.5\% samples was observed in the first 4 months after production; it then remained constant at a level of $\sim$ 65 tracks cm$^{-2}$ d$^{-1}$ (at the $^{210}$Po peak). No  significant increase was observed for the PbCa 0.07\% and PbCa 0.03\% over a period of more than 1 year for all energy intervals; the radioactivity remained at the level of 1-3 tracks  cm$^{-2}$ d$^{-1}$ at the energy peak corresponding to $^{210}$Po decay. Co-lamination with thin layers of pure lead to protect the emulsions from $\alpha$-radioactivity  was found to be technically difficult and hence not to be a viable solution. \par

An increase of the fog in  emulsions packed in vacuum was observed. This effect is negligible for the bricks packed in air rather than in vacuum, as currently done for  OPERA  emulsions.The packing technique for the bricks was decided accordingly to this observation. \par

The presence of  $\alpha$- and $\beta$-rays from the lead radioactivity may affect the track reconstruction in the emulsions. However, it was found that the spurious tracks (or more precisely base tracks) can be removed by a simple quality cut. The reconstruction efficiency for some types of event has to be studied and further tests are in progress. The visual inspection of candidate events is envisaged to further improve the situation and eliminate problems.\par
 
 On the basis of the results obtained it was decided to use for OPERA the PbCa 0.03\% alloy, which has good mechanical properties and is safe with respect to $\alpha$-radioactivity and chemical effects.

\section {Acknowledgements}

We acknolewdge the  cooperation with the group of Prof. V. Palmieri of INFN-LNL for discussions, advices and preparation of some etched samples.  We thank the Health Physics Group of the Physics Department of Bologna, in particular Dr. M. P. Morigi, and Dr. M. Bettuzzi for providing the SBSiD detector equipment and for helpful suggestions. We are grateful to Dr. Z. Sahnoun for her collaboration.

We warmly acknowledge funding from our national agencies: 
{\it Fonds de la Recherche Scientifique - FNRS et Institut Interuniversitaire des Sciences Nucl\'eaires} for Belgium, 
MoSES for Croatia, 
IN2P3-CNRS for France, 
BMBF for Germany, 
INFN for Italy, 
the {\it Japan Society for the Promotion of Science} (JSPS), 
the {\it Ministry of Education, Culture, Sports, Science and Technology} (MEXT) and the {\it Promotion and Mutual Aid Corporation for Private Schools of Japan} for Japan, 
SNF and ETHZ for Switzerland, 
the {\it Russian Foundation for Basic Research} grants 06-02-16864-a and 08-02-91005-CERN-a for Russia,
the {\it Korea Research Foundation} (KRF-2007-313-C00161) for Korea.

We thank INFN for providing fellowships and grants to non Italian researchers. We thank ILIAS-TARI for access to the LNGS research infrastructure and for the financial support through EU contracts P2006-01-LNGS and P2006-16-LNGS.

We are finally indebted to our technical collaborators for the excellent quality of their work over many years of design, prototyping and construction of the detector and of its facilities.

\bibliographystyle{plain}

\end{document}